\documentclass{article}

\usepackage{arxiv}

\usepackage[utf8]{inputenc} 
\usepackage[T1]{fontenc}    
\usepackage{hyperref}       
\usepackage{url}            
\usepackage{booktabs}       
\usepackage{amsfonts}       
\usepackage{nicefrac}       
\usepackage{microtype}      
\usepackage{lipsum}
\usepackage{graphicx}
\usepackage{float}
\usepackage{subfigure}
\usepackage{cite}
\usepackage[version=4]{mhchem}
\graphicspath{ {./images/} }

\title{Design of Programmable Temperature Platform and its  Pyroelectrocatalytic applications}

\author{
 Xiechao Hu \\
  School of Computer Science\\
  Northwestern Polytechnical University\\
  Xi’an 710129, China \\
  \texttt{13572929906@163.com} \\
   \And
 Chengxi Hu \\
  School of Materials Engineering\\
  Xi’an Aeronautical Institute\\
  Xi’an 710077, China \\
  \texttt{huchengxi@163.com} \\
  \And
 Tieyan Guo \\
  School of Materials Engineering\\
  Xi’an Aeronautical Institute\\
 Xi’an 710077, China \\
     \And
 Zhi Yao \\
  School of Materials Engineering\\
  Xi’an Aeronautical Institute\\
 Xi’an 710077, China \\
    \And
 Yang Yang \\
 School of physics and information technology\\
 Shaanxi Normal University\\
Xi’an,710062, China\\
    \And
 Ze Qing Guo \\
 Key Laboratory of Functional Materials and Devices for Informatics of Anhui Educational Institutions\\
Fuyang Normal University\\
Fuyang 236037, China \\
    \And
Amanda Ekeminiabasi Williams \\
 School of Electronic Information and Artificial Intelligence\\
Shaanxi University of  Science  and 
Technology\\
Xi’an, 710021, China \\
}

\begin{document}
\maketitle
\begin{abstract}
The Si based \ce{TiO2} thin films were prepared via the combination both of Sol-Gel and Spin-Coating method. The films were sintered at 850℃ for half an hour, and the resulting films were characterized by X-ray diffraction (XRD) and scanning electron microscopy (SEM) for their phase composition and microstructure. It was found that the films contained silicon, anatase phase, and unknown impurities. There were holes and micro-cracks on the surface of \ce{TiO2} films. A program-controllable hot-cold test chamber was successfully developed and used to study the catalytic performance of the thin film for the first time. The results showed that \ce{TiO2} thin films had the ability to degrade RhB dyes. The highest degradation rate of RhB achieved 37\% after 48 cold-hot cycles. Our design and the experimental results presented in this paper strongly highlight the bright prospects of the thermoelectric properties of \ce{TiO2} and water environmental disinfection applications.
\end{abstract}


\section{Introduction}
Due to the development of industrialization and the resulting large-scale population, the energy shortages and environmental pollution are two major issues that deserve attention in today’s world\cite{r1,r2,r3,r4,r5,r6,r7,r8,r9,r10}. Pyroelectrocatalytic research is a very active and frontier field involving the processes that utilize light energy to facilitate chemical reactions\cite{r11,r12,r13,r14,r15,r16,r17,r18,r19,r20,r21}. The technology is expected to be used in areas such as environmental governance, energy production and organic synthesis. As an environmental friendly and efficient pollution control technology, titanium dioxide catalysis has received wide attention and research in recent years\cite{r22,r23,r24,r25,r26,r27,r28,r29,r30,r31,r32,r33,r34,r35,r36,r37}. The current research status mainly includes the following aspects: Firstly, Researchers have conducted in-depth research on the mechanism of titanium dioxide catalysis, including the mechanism analysis of optical absorption, electron transfer, active site and other aspects, which provides a theoretical basis for further optimizing the catalysis effect. Secondly, in order to improve the catalytic performance of titanium dioxide, researchers continuously optimize the structure and performance of the catalyst through modification methods, such as metal doping, nanostructure design and other means. Thirdly, titanium dioxide catalysis has been widely used in organic wastewater treatment, air purification, catalysis catalytic hydrogen production and other fields, it’s related application research has also made a lot of progress. For example, more efficient wastewater treatment effects can be achieved. Fourthly, some research institutions and enterprises have begun to explore the industrial application of titanium dioxide catalysis technology, such as the increasing number of application cases in the field of water treatment, which provides strong support for the promotion and application of this technology. In general, the research of titanium dioxide catalysis has made some progress in theoretical exploration, experimental research and application. In the future, it is necessary to increase the research efforts, constantly explore new improvement methods, application fields and promote the further development and application of titanium dioxide catalysis technology. \\

However, the current catalysts have the following disadvantages: Firstly, in the long-term use process, the catalysts is easily affected by light, redox and other factors which leads to the decline or failure of its catalytic activity. Secondly, most catalysts can only absorb light energy in the range of ultraviolet or visible light and cannot make full use of broad spectrum light sources such as sunlight which limits their application range and efficiency. Thirdly, the preparation cost of some efficient catalysts is high which limits their popularization in large-scale applications. Fourthly, some catalysts are prone to contamination or inactivation in the reaction process and regeneration is difficult which affects their long-term stability and recycling performance\cite{r37,r38,r39}.\\

Furthermore, most of the current catalysts are powder, although powder catalysts have many advantages in the field of catalysis, they also have obvious disadvantages. First of all, it is difficult to effectively recycle and reuse the powder catalysts scattered into the reaction solution, leading to the failure and waste of the catalyst. Powder catalysts usually require UV light as an excitation light source and the use of UV light can increase energy costs and safety risks. Besides, powder catalysts can easily degrade or fail under high temperature or long time light which affects its lasting catalytic effect.\\

Compared with powder catalyst, thin film catalysts have absolute advantages as follows: 1. The thin film catalyst has a high specific surface area, which can improve the rate of catalytic reaction and effectively promote the catalytic process. 2. Strong light absorption capacity: after special treatment of the surface of the thin film catalyst, it can effectively absorb light energy and improve the catalytic efficiency. 3.Film catalyst can realize the controllability of catalytic reaction by adjusting its thickness, composition and other parameters. 4. The thin film catalyst has good stability and recycling that can achieve long-term stable catalytic reaction. 5. Thin-film catalysts can be easily loaded on different substrates, making them have a wider range of applications.\\

Therefore, it is interesting to explore the preparation of \ce{TiO2} thin film and to investigate the pyroelectrocatalytic reactivity, which there is no literature discussed up to date. In this paper, we reported that the eco-friendly \ce{TiO2} thin film sintered at 850 ℃ using the Sol-Gel and Spin coating method, displaying the highly catalytic efficiency for decomposition organic pollutants.\\

In addition, traditional alternating temperature chambers usually have high energy consumption, increased costs, high maintenance cost, harmful gas emissions and inconvenient temperature control. In order to simulate the effect of the diurnal temperature difference on the pyroelectrocatalysis of the thin films, we designed a programmable temperature control platform with many advantages. It can be set in a variety of temperature control modes according to the needs of the experiment, such as constant temperature, temperature rise and cooling, etc. It is easy to operate, and the temperature can be controlled by presetting programs or setting parameters. In this paper, we used it to simulate the effect of diurnal temperature differences on the catalytic performance of the \ce{TiO2} thin film\cite{r35,r36,r37,r38,r39,r40,r41}

\section{Experimental procedure}
\subsection{Preparation of the \ce{TiO2} Sol}
(1) Use decontamination powder to remove container pollution, take a certain amount of butyl titanate (\ce{C16H36O4Ti}), absolute ethanol (\ce{C2H6O}), mix at 11:17 volume ratio and stir magnetically for 10 min, to obtain solution A.\\

(2) Take a certain amount of water, ice-cold acetic acid, nitric acid, mix according to the volume ratio of 17:8:2, add an appropriate amount of sodium 1-octane sulfonate (\ce{C8H19NaO4S}) and magnetic stirring for 10 min, to obtain solution B.\\

(3) Drop solution A into solution B (2 drops / s), and stir the solution continuously for 30 min to obtain a uniform light yellow transparent sol.\\

The preparation process of \ce{TiO2} Sol is shown in Figure 1.\\

\begin{figure}[htb]  

      \centering  

      \includegraphics[width=0.9\linewidth]{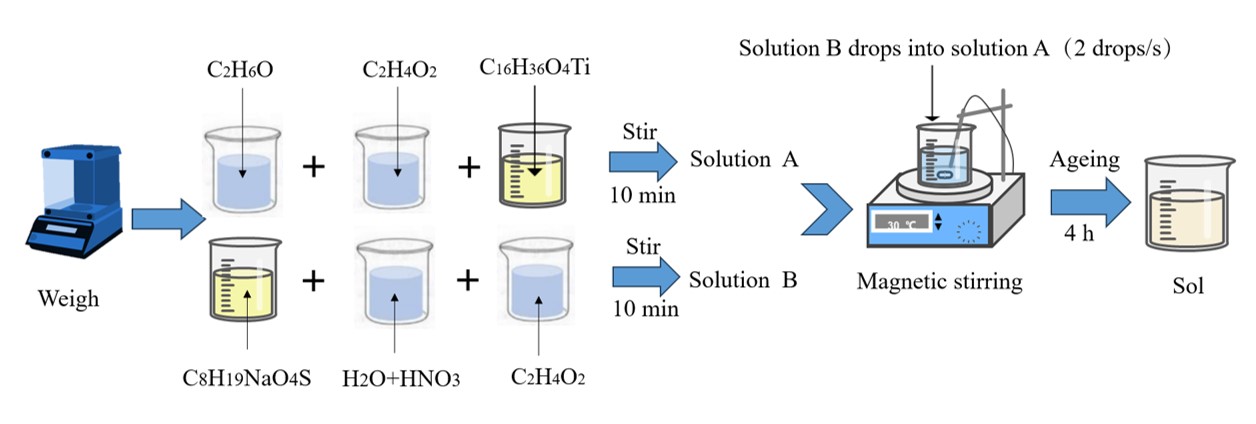} 

      \caption{Flow chat of preparation \ce{TiO2} colloid}

      \label{figure 1} 

\end{figure}

\subsection{Preparation of \ce{TiO2} thin films}
(1) The Si substrate was sonicated in deionized water for 10 min, and then placed in absolute ethanol for 10 min. Before spinning the coating, soak the silicon water in absolute ethanol and dry it in a drying box.\\

(2) Place the Si substrate on the rotary coating machine with tweezers, drop the \ce{TiO2} sol in the center of the substrate, turn on the power supply of the rotary coating machine, set the low speed to 350 rpm, the high speed to 3000 rpm, and the rotary coating time is 10s.\\

(3) After the rotary coating, allow the substrate to dry for a period of time, then put into a drying box, and dry at 170℃ for 20 min. After cooling to room temperature, the next coating, a total of 5 times.\\

(4) The spin-coated substrate was put into the furnace and annealed at 850℃ for 90 min to finally obtain the \ce{TiO2} film based on Si.
The preparation process of \ce{TiO2} thin films is shown in Figure 2.\\

\begin{figure}[htb]  

      \centering  

      \includegraphics[width=0.9\linewidth]{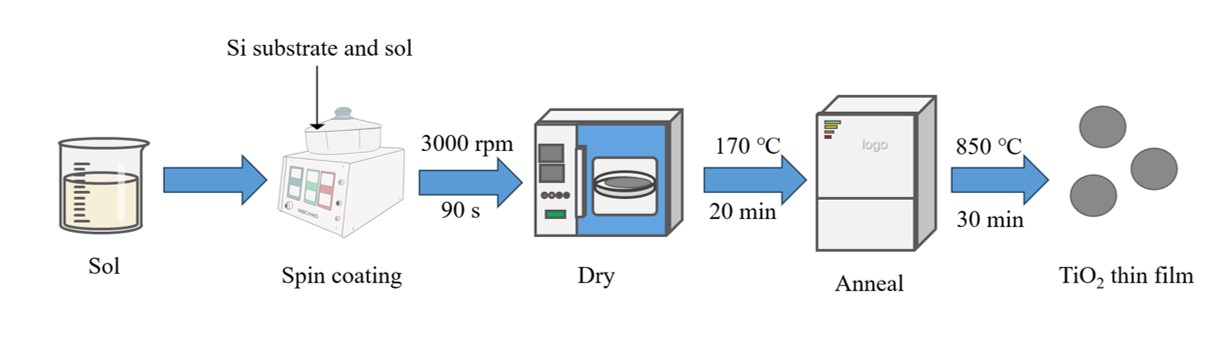} 

      \caption{Flow chat of \ce{TiO2} thin film prepared by spin coating}

      \label{figure 2} 

\end{figure}

\subsection{The micro-structural characteristics of the thin films.}
The XRD patterns obtained on D/max-2550V/PC X-ray diffractometer (Rigaku, Japan) using Cu k$\alpha$
 radiation at a scan rate 2 s$^{-1}$ were used to determine the identity of crystalline phases. The morphology of the prepared samples was characterized by high-resolution transmission electron microscopy (JEM-2100, Japan) and a scanning electron microscopy (SEM, FEI-quanta200, USA). The X-ray Photoelectron Spectroscopy (XPS) measurements were performed on an AXIS ULTRA spectrometer (Kratos  Analytical Ltd, Japan) with a monochromatic Al. 

\subsection{Design of programmable temperature control system}
The programmable temperature control table system includes hardware and software. The hardware consists of controller, thermometer, relays, wires and semiconductor refrigeration sheets, as show in Figure 3. The software is written in C language.

\begin{figure}[htb]  

      \centering  

      \includegraphics[width=0.9\linewidth]{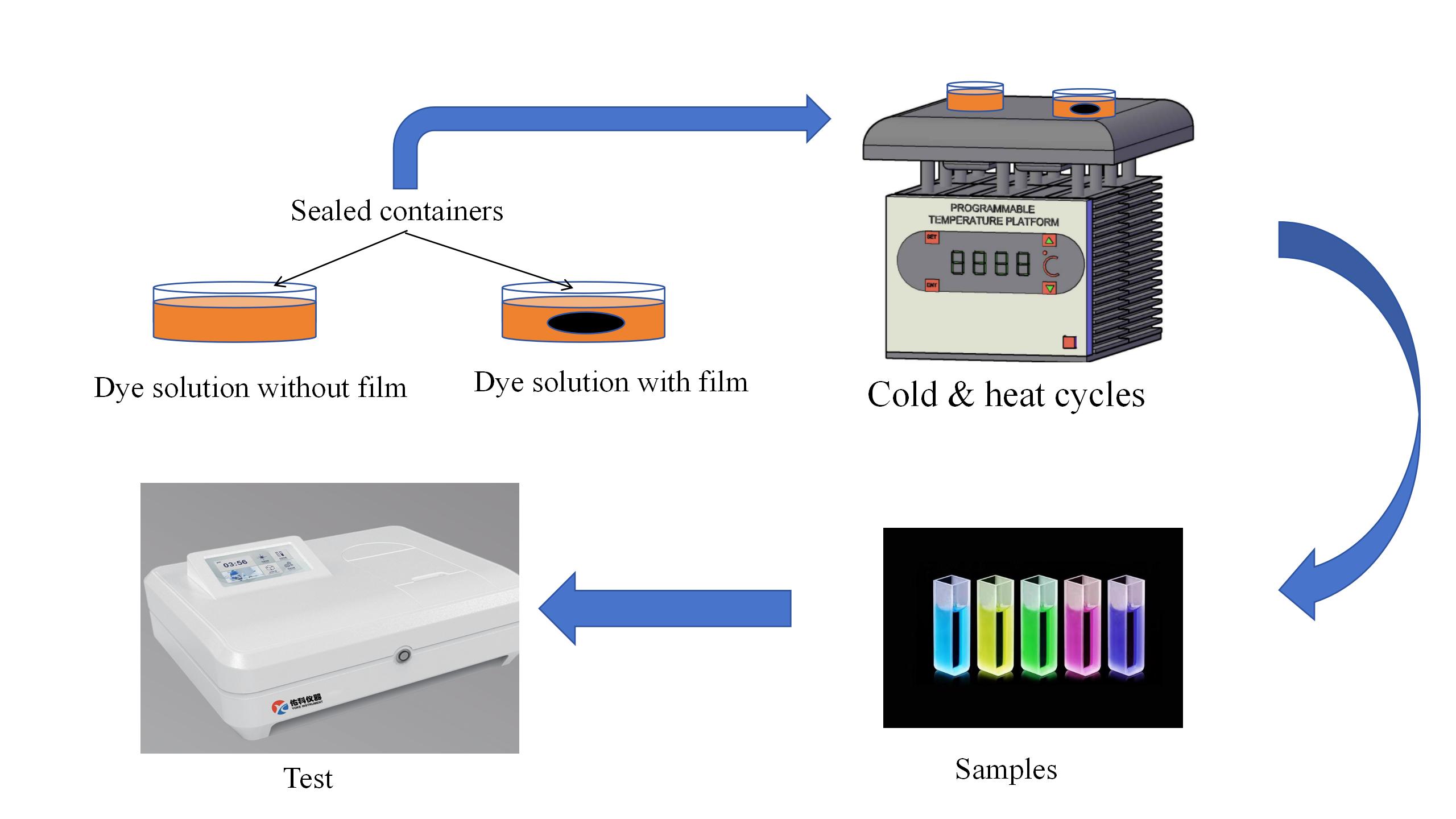} 

      \caption{Flow chat of \ce{TiO2} thin film piezoelectrocatalytic degradation}

      \label{figure 3} 

\end{figure}

\subsection{Catalytic degradation experimental}
The \ce{TiO2} film was placed into 5 mg/L of RhB, MO, MB solution, and the piezoelectric catalytic degradation experiment of the dye solution was performed under the condition of ultrasonic vibration. The specific experimental steps are described as follows:\\

(1)Accurately weigh and dispense 5 mg/L of RhB, MO and MB solution  respectively with electronic analytical balance. Take two copies of 100 ml of each dye solution, one without any treatment, labeled with A, and one in \ce{TiO2} film, labeled with B.\\

(2) The two solutions were sealed to avoid the influence of water vapor on the experiment. The sealed solution is subjected to ultrasonic vibration treatment. Light must be shaded during the experiment.\\

(3) The water temperature shall not exceed 50℃ during ultrasound treatment. Samples were drawn at 10-min intervals with needle tubes.\\

(4) After the sampling process, the absorbance spectrum of the sample solution is tested by using a spectrophotometer. The height of the sample solution shall exceed 1 / 3 of the height of the cuvette and shall not exceed 2 / 3.\\

(5) Calculate the degradation rate of the sample solution. The solution degradation rate was calculated by the following formula.\\

\begin{equation}
D = \left( 1 - \frac{A_t}{A_0} \right) \times 100\% \tag{1}
\end{equation} \\

Where $D$ was the degradation rate of dye solution, $A_0$ and $A_t$ were the absorbance of the solution at the initial time and at time $t$.
\\

The experimental process of \ce{TiO2} thin film piezoelectrocatalytic degradation is shown in Figure 3.

\section{Results and discussion}
Figure 4 shows the XRD patterns of \ce{TiO2} thin films. It could be seen that all the main diffraction peaks can be well indexed to the standard patterns of Si (PDF \#8 0-0018) and \ce{TiO2} (PDF \# 88-1173), indicating that the solid thin film was formed at monocrystalline silicon as a anatase phase. A very little rutile  secondary phase appeared along with the main phase anatase in all samples, which is similar to the results of J S et al[10]. 
\begin{figure}[htb]  

      \centering  

      \includegraphics[width=1.0\linewidth]{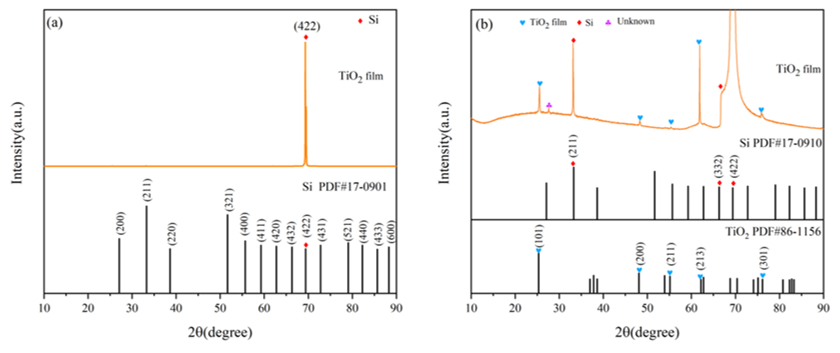} 

      \caption{XRD patterns of \ce{TiO2} thin film prepared by combination of Sol-gel and spin-coating method}

      \label{figure 4} 

\end{figure}

From Figure 4(a), we can see that there are two super-strong peaks around 69°, corresponding to crystal face (400) and (301). All diffraction peaks correspond to the standard card PDF \# 80-0018 for Si and the standard card PDF \# 88-1173 for \ce{TiO2}. Figure 3(b) is a local magnification of XRD of \ce{TiO2} films. The diffraction peaks at 27.717, 36.303, 39.599, 41.537, 44.508, 54.801, 57.246 and 69.700 belong to (110), (100), (111), (210), (211), (220) and (301), corresponding to \ce{TiO2}(PDF \# 88-1173), and diffraction peaks at (052), (310) crystal surfaces of \ce{TiO2} PDF \# 78.046.\\

Figure 5 shows the SEM images and EDS mapping analysis of the thin film. Both of the film without and with Ag doped show the highly flatness and uniformity, And it can clearly observe the silver particles with an average particle size of 300 nm as shown in Figure 5(b-d). \\

\begin{figure}[htb]  

      \centering  

      \includegraphics[width=0.8\linewidth]{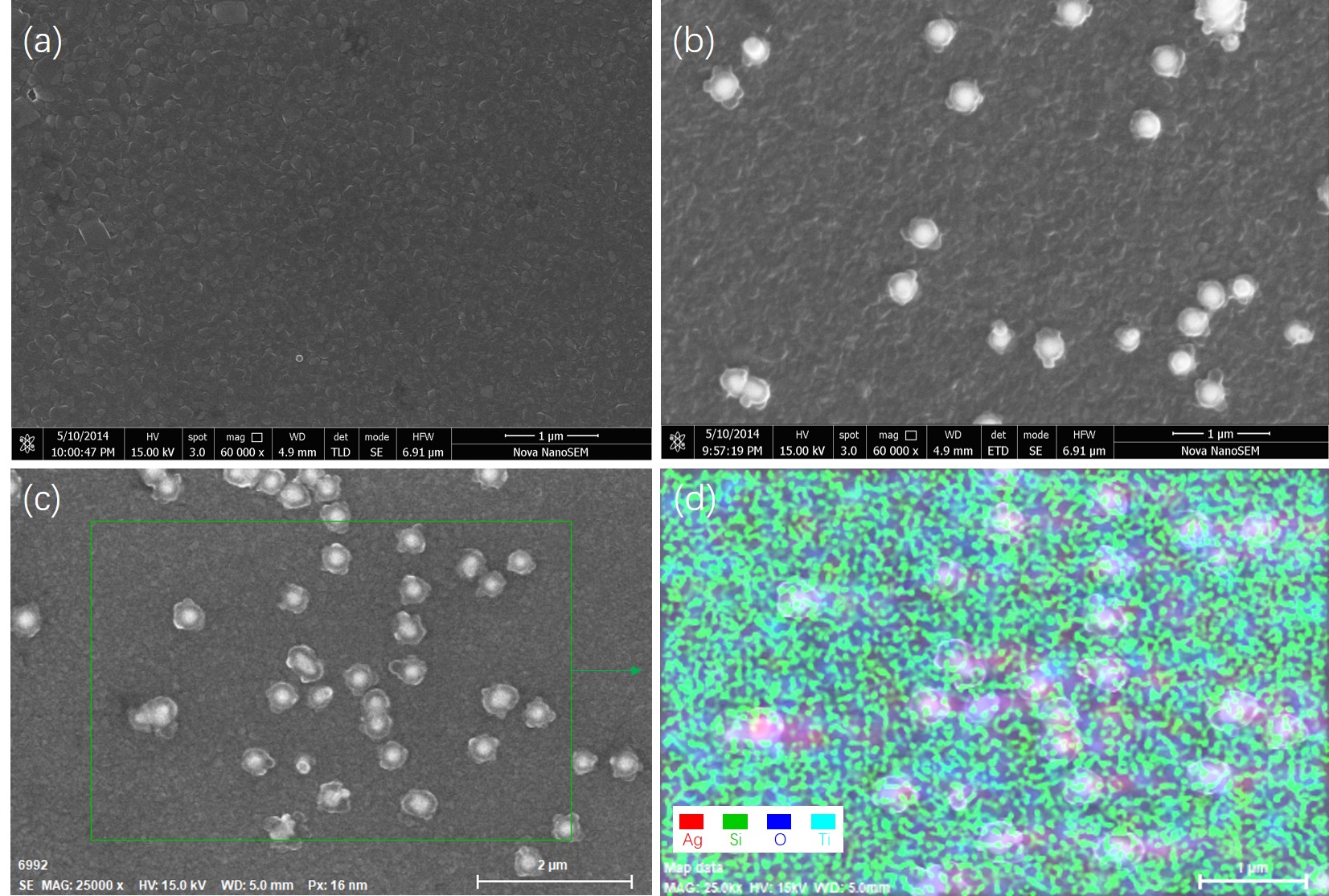} 

      \caption{SEM images of \ce{TiO2} thin film prepared by  combination of Sol-gel and spin-coating method:}

      \label{figure 5} 

\end{figure}

\begin{figure}[t]
\centering  
\renewcommand{\thesubfigure}{(\Alph{subfigure})} 
\subfigure[by different thin film (a) initial; (b) without film after 12 cycles; (c) with \ce{TiO2} after 12 cycles; (d) with \ce{TiO2} doped Ag after 12 cycles.]{
\label{Fig.sub.1}
\includegraphics[width= 7.1cm,height = 6.0cm]{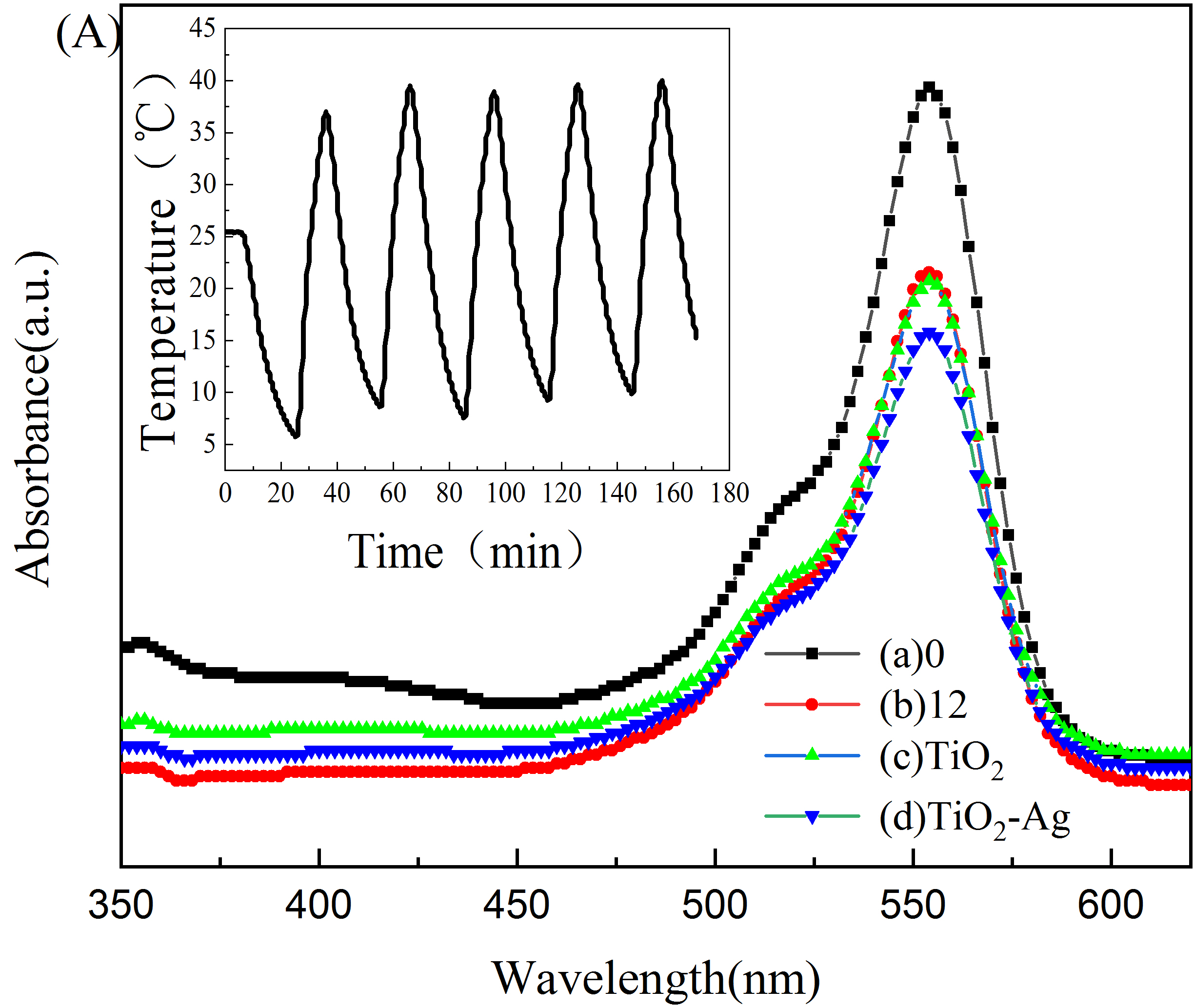}}\subfigure[by \ce{TiO2} doped Ag film.]{
\label{Fig.sub.2}
\includegraphics[width= 7.3cm,height = 6.0cm]{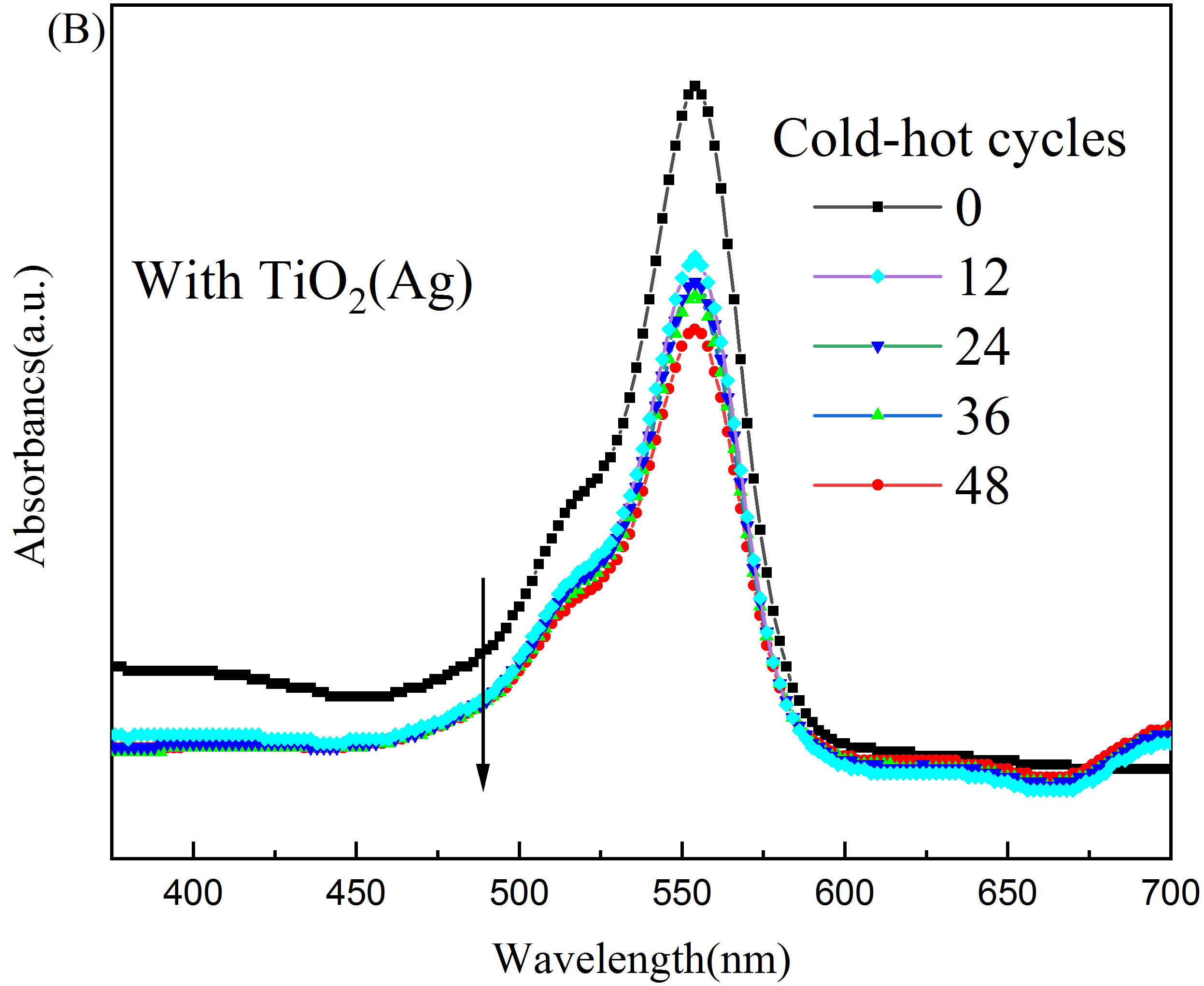}}
\caption{UV-vis spectra of RhB solutions experiencing different cold-hot cycles.The inset in (A) is the setup temperature for cold-hot cycles.}
\label{1}
\end{figure}

For the \ce{TiO2} film, the degradation of RhB were evaluated using a Silicon-based film with a diameter of 4 inches placed in a 50-mL RhB solution contained in a 100-mL closed flat-bottomed metal utensil as shown in Fig. 3. The pyro-driven catalytic activity was evaluated by the decomposition of RhB dye. Fig. 6 shows the UV-vis spectrum of RhB dye solution with and without different films after experiencing different cold–hot cycles, where the setup temperature for each cold–hot cycle is shown in Fig. 6(A). It is clearly that absorption peak of the solution with both \ce{TiO2} and \ce{TiO2} doped Ag film is lower than that without film. Moreover, absorption peak of the solution with \ce{TiO2} doped Ag film is lower than that with \ce{TiO2} film. This indicates that silver doping is beneficial to improve the catalytic performance of \ce{TiO2}. The maximum absorption peak (~554 nm) of RhB solution with \ce{TiO2} doped Ag film decreased continuously with increasing cold–hot cycles. To easily observe the decomposition process, Fig.7 shows the decomposition ratios (D) of RhB solution ($D = \left( 1 - \frac{C_0}{C} \right) \times 100\%$), where $C_0$ and $C$ are the initial absorbance and the measured absorbance of maximum peak at the different cold-hot cycles, respectively) and the corresponding optical graph of RhB solution after experiencing different cold-hot cycles. Clearly, the color of RhB solution fades with the cold-hot cycles , indicating that RhB decomposed with increasing cycles.

\begin{figure}[htb]  

      \centering  

      \includegraphics[width=0.6\linewidth]{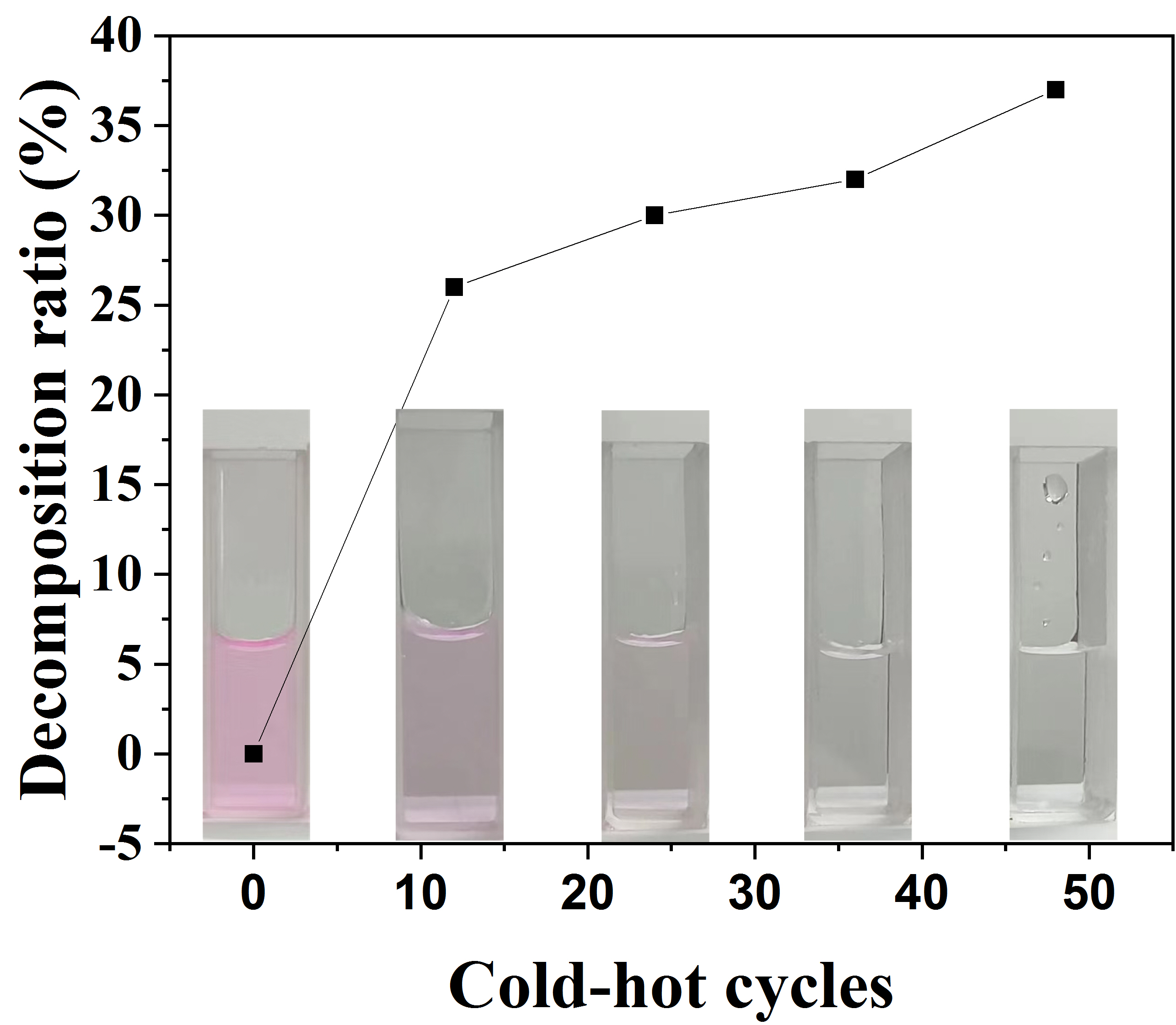} 

      \caption{Decomposition ratios of \ce{TiO2} film after experiencing different cold-hot cycles. The bottom is a corresponding optical photo of RhB solution after experiencing various cold-hot cycles.}

      \label{figure 7} 

\end{figure}

\section{Conclusions}
Highly flatness and homogeneity \ce{TiO2} thin films were prepared onto Si substrate base by combination both of sol-gel and spin coating. The films were sintered at 850 °C for half an hour, and the resulting films were characterized by X-ray diffraction (XRD) and scanning electron microscopy (SEM) for their phase composition and microstructure. It was found that the films contained silicon, anatase phase, and unknown impurities. The surface of the \ce{TiO2} thin film was relatively flat, and the grain distribution was uniform, with microcracks present. The dye degradation experiment was conducted by performing cold and hot cycle tests, and the absorbance of the dye solution was measured using a spectrophotometer. The results showed that both of with and without Ag doped \ce{TiO2} thin films had the ability to degrade RhB. The highest degradation rate of RhB solution was 37\% after experiencing 48 cycles of cold and hot cycles. The pyro-driven catalytic properties of \ce{TiO2} thin films can effectively degrade water pollutants and can be applied in related fields of environmental governance.

\section*{Acknowledgements}
This work was supported by Scientific Research Foundation of Education Department of Province Anhui (2024AH051453, KJ2020A0528).School-level youth talent program of Fuyang Normal University (RCXM202214).Horizontal Cooperation Project of Fuyang Municipal Government-Fuyang Normal University (FMDI202402).

\bibliographystyle{unsrt}
\bibliography{references}
\end{document}